\documentclass[12pt]{article}
\usepackage[backend=biber, style=apa, natbib=true]{biblatex}          
\usepackage{newtxtext}                
\usepackage{newtxmath}              
\usepackage{geometry}                
\usepackage{graphicx,xspace,amsmath}
\usepackage{bm}
\usepackage{xcolor}
\usepackage{hyperref}
\usepackage{soul}
\usepackage{dcolumn}
\usepackage{booktabs}
\usepackage[normalem]{ulem}
\newcommand{\ie}{\emph{i.e.}\xspace}

\bibliography{DaPo.bib}	
\begin{document}
\author{
	\\ 
	\\
	\rm Francesco Dotto\\
	\it Universit\`a Roma Tre, Italy\\
	\texttt{francesco.dotto@uniroma3.it}
	\\
	\\ \rm
	Richard D. Gill\\
	\it Leiden University, The Netherlands\\
	\texttt{gill@math.leidenuniv.nl}
	\\ 
	\\
	\rm  Julia Mortera\thanks{corresponding author}\\
	\it Universit\`a  Roma Tre, Italy\\
	\texttt{julia.mortera@uniroma3.it}
}

\title{Statistical Analyses in the case of an Italian nurse accused of murdering patients}

\date{}

\maketitle

\begin{abstract}
	Suspicions about medical murder sometimes arise due to a surprising or unexpected series of events, such as an apparently unusual number of deaths among patients under the care of a particular nurse. But also a single disturbing event might trigger suspicion about a particular nurse, and this might then lead to investigation of events which happened when she was thought to be present. In either case, there is a statistical challenge of distinguishing event clusters that arise from criminal acts from those that arise coincidentally from other causes. We show that an apparently striking association between a nurse's presence and a high rate of deaths in a hospital ward can easily be completely spurious. In short: in a medium-care hospital ward where many patients are suffering terminal illnesses, and deaths are frequent, most deaths occur in the morning. Most nurses are on duty in the morning, too. There are less deaths in the afternoon, and even less at night; correspondingly, less nurses are on duty in the afternoon, even less during the night. Consequently, a full time nurse works the most hours when the most deaths occur. The death rate is higher when she is present than when she is absent.
\end{abstract}

\noindent {\small {\em Some key words:} 
	clusters of unexplained events; causation versus association; GLM; health care serial killers; investigative bias.}

\section{Introduction}
\label{sec:intro}
This paper discusses the case of an Italian nurse, Daniela Poggiali (DP), who was an experienced  nurse  at the ``Umberto I'' hospital in Lugo, in the province of Ravenna, where she worked  between  9/4/2012 and 8/4/2014.  DP was initially accused of  the murder of a 78-year-old patient, Rosa Calderoni (RC) on the 8$^{\mbox {th}}$ of April 2014. Following police investigation and trial, a judge concluded that she  was  guilty of having injected patient Rosa Calderoni with a lethal dose of potassium chloride and on 11/3/2016 sentenced her to life imprisonment.  Her employment had been terminated from the day after the death of her patient RC. The main evidence in this trial was based two arguments. First,  the fact that there had been an upset about potassium chloride which had gone missing and later turned up somewhere where it shouldn't have been. Second, the two witnesses for the prosecution, Prof.~Franco Tagliaro (Univ.~Verona) and Prof.~Rocco Micciolo (Univ.~Trento)  claimed that since DP had a significantly higher death rate in her presence, this could not be due to chance and implied ``la pratica seriale dell'omicidio dei pazienti'' (the workings of a serial patient killer). 

 A further trial independently took place on the 15th of December 2020, at which
 Daniela Poggiali was sentenced to thirty years in prison, accused of murdering another
 one of her patients, this time 95-year-old Massimo Montanari (MM), the former employer
 of her partner, who died on March 12, 2014 (a month before the death of RC). The
 prosecution considered the fact that DP was overheard in an altercation with MM in
 2009, 5 years before his death, as evidence against her. The main evidence in this trial
 was based solely on the statistical evidence of the two witnesses for the prosecution,
 again Tagliaro and Micciolo.
 
\subsection{Prosecution's hypothesis}
Here we briefly show the prosecution's hypothesis based on the analysis of death rate. The two experts, Tagliaro and Micciolo, which will be referred to as TM from now on, considered  daily and weekly death rates, showing that DP had the highest death  rates when on duty than the other nurses. 
 TM claim, in their report, that ``the most powerful analysis'' of the death rate of the Lugo hospital can be summarized by the following table

\begin{table}[ht]
	\centering
	\begin{tabular}{c|rcccccrr}
		\hline
		&      \multicolumn{1}{c}{same} &   opposite &      total &      hours & \multicolumn{2}{c}{mortality rate}           &   \multicolumn{1}{c}{relative} &   absolute \\
		nurse	&      \multicolumn{1}{c}{zone}  &       zone &     deaths &    on duty &  same         &     opposite       &      \multicolumn{1}{c}{risk}  &      \multicolumn{1}{c}{risk}  \\
		\hline
		N.1 &         68 &         58 &        126 &       3686 &       0.54 &       0.46 &       1.17 &       0.08 \\
		N.2 &         51 &         68 &        119 &       3545 &       0.43 &       0.57 &       0.75 &      -0.14 \\
		N.3 &         64 &         60 &        124 &       3554 &       0.52 &       0.48 &       1.07 &       0.03 \\
		N.4 &         70 &         53 &        123 &       3535 &       0.57 &       0.43 &       1.32 &       0.14 \\
		N.5 &         64 &         41 &        105 &       3625 &       0.61 &       0.39 &       1.56 &       0.22 \\
		N.6 &         43 &         65 &        108 &       3532 &       0.40 &       0.60 &       0.66 &      -0.20 \\
		DP &        139 &         52 &        191 &       3577 &       0.73 &       0.27 &       2.67 &       0.46 \\
		N.8 &         60 &         44 &        104 &       3710 &       0.58 &       0.42 &       1.36 &       0.15 \\
		N.9 &         66 &         53 &        119 &       3741 &       0.55 &       0.45 &       1.25 &       0.11 \\
	\end{tabular}  	
	\caption{Table representing  TM's analysis of mortality rates showing the relative and absolute risk in the same and in the opposite zone when the nurses are on duty.}
	\label{tab:micciolo}
\end{table}

\begin{figure}
\begin{center}
\caption{A pictorial representation of the structure of the Lugo hospital ward.}
\label{fig:str}
\includegraphics[scale=.7]{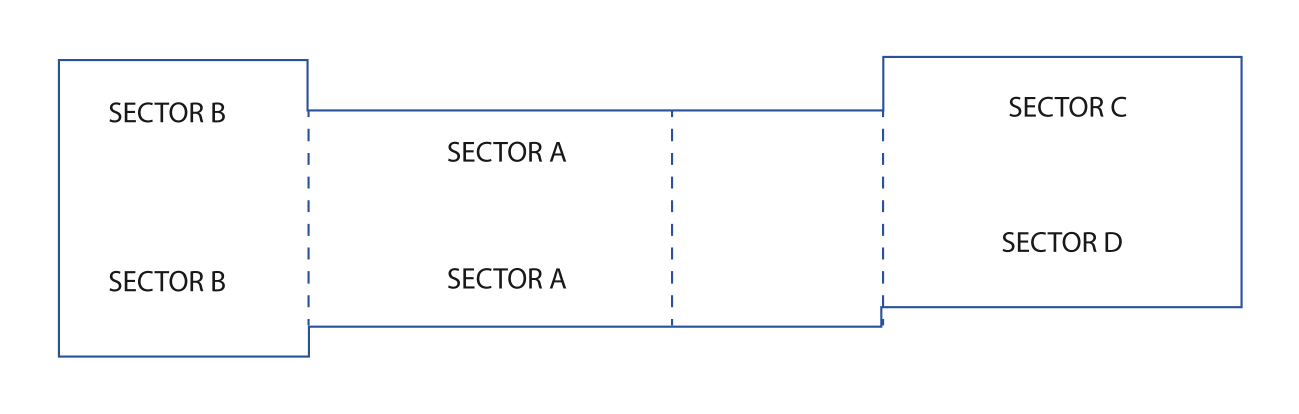}
\end{center}
 \end{figure}

 Table \ref{tab:micciolo} needs to be interpreted as follows. TM divide the hospital ward in two \textit{zones} made up of the two contiguous sectors, specifically sectors A + B and sectors C + D. If a nurse is working in sector A while a death occurs in sector B, the death is designated as being ``in the same zone''. A death in C while she works in A is called ``in the opposite zone''. Figure \ref{fig:str} shows the structure of the Lugo hospital ward.
 
 Then, for each shift, they compute the number of deaths recorded when a given nurse is on duty in a given zone and the number of deaths recorded in the opposite zone.  DP has the highest patient mortality rate in the zone where she is on duty  compared to the mortality rate in the opposite zone  and thus the highest relative and absolute risk. This fact led TM to infer a causal effect between DP's presence  and the increase in deaths. For the sake of brevity we reproduce their computations in Table \ref{tab:micciolo}  only for those nurses that work a similar number of hours as DP. 

 We believe that there is a rather simple innocent explanation for a substantially increased mortality rate. DP is a full time and fully qualified nurse. There are three shifts every day, but the morning shift has many more nurses on duty than the afternoon, and in the evening shift there are even less. A nurse like DP works many more morning shifts than afternoon shifts, and even less night shifts. At the same time, most deaths occur (or at least, are registered) during the morning shift, for well understood biological reasons. Patients like those in the Lugo wards do not die in their sleep -- they die in the morning, as the body's organs start working at full power. Failing systems break down at this moment, not in the night when everything is in something like a hibernation mode. There are less deaths in the afternoon and evening, and even less in the night. Hence a nurse like DP works many more morning shifts than afternoon or night shifts, and there are more deaths during her presence than during her absence.

As we will see, there are yet more factors which make such raw mortality rates difficult to evaluate. Firstly, the registered times of deaths contain many anomalies due to the fact that a death has to be registered by a qualified doctor who reports \emph{the time at which they sign the death certificate}. Many deaths are registered in the times of hand-over between one shift and the next, during which time nurses are present who work both shifts. A nurse who arrives well in time for the hand-over and only leaves well after the next hand-over is complete therefore experiences many more deaths during her presence than one who keeps her shifts as short as possible.

Furthermore, as we will later see, DP had been moved from one ward to another several times in the months just before the death of RC. Hospital authorities were in fact already suspicious of her, and had already found out that recently, death rates when she was on duty were twice what they were when she was not on duty. There was already gossip about her among the nurses and other supporting personel associating her to recent alleged thefts of patients' jewelry, and there had been an upset about potassium chloride which had gone missing and later turned up somewhere where it shouldn't have been. Tagliaro became formally involved in the case at the end of the year, immediately recruiting Micciolo to support him on the statistical analyses; the two of them spent two months (December 2014 and January 2015) writing their statistical report.

 In other similar cases, nurses become the focus of attention as they are an odd-ball. For example,  Lucia de B.\ attracted attention because of her colourful personality and thanks to gossip about her personal history. Later, while being interrogated by police and when appearing in court, her fashion choices were interpreted, for instance by an FBI profiler giving evidence for the prosecution, and also in some of the media, as brazen efforts to win sympathy. Certainly, Daniela Poggiali was a nurse with a strong and colourful personality and a sharp tongue. Like Lucia, she stood out in the crowd. She had earlier complained to her colleagues that RC was a difficult patient.

Expanding on the issues raised above, we structure the paper as follows.  In Section \ref{sec:data}, we outline the data used in 
 \ref{sec:time} and \ref{sec:shifts}. We analyse the age distribution of the deceased  patients and the distribution of the sectors in which DP is on duty. In Sections \ref{sec:time} and \ref{sec:shifts}  we present our analysis of the mortality rate in the hospital sectors by time and by nurses' shifts. We display the number of nurses on duty and the corresponding number of deaths for DP and the other nurses in Section \ref{sec:service}. Section \ref{sec:bayes} gives an overview of the misinterpretation of clusters of mortality events and the use of Bayes' rule in these cases. In Section \ref{sec:GLM} we present a new analysis of the data using generalised linear models  and in Section  \ref{sec:pred} we illustrate the pitfalls in the analysis of the potassium data. Finally, in Section \ref{sec:pred}, we also show that \emph{statistical} computations made in a second report by the pathologist Tagliaro on the amount of potassium in the vitreous fluid of the eye of the deceased are incorrect and in no way incriminating for DP. 
Concluding remarks are given in Section \ref{sec:conc}.

\subsection{History of the trials}

On the 11$^{\mbox {th}}$ of  March 2016, Daniela Poggiali was sentenced by the Ravenna Court of Assizes to life imprisonment.  The judge declared that she  was  guilty of having injected patient Rosa Calderoni with a lethal dose of potassium chloride. Richard Gill and Julia Mortera had testified on 18/12/2015 for the defence in this trial, called to give a critique of a report written by two ``statistical experts'' TM.  About our testimony the judge wrote:
\textit{As a matter of fact no effective technical element has been brought by the defence consultants, who, remaining on theoretical issues, have not undermined the valuable technical report of the prosecution.} Our testimony, as we will show,  was far from being theoretical. Here  we will also illustrate the pitfalls and misinterpretation of the data in TM's report that was used against DP. In Section \ref{sec:pred} we also show that \emph{statistical} computations made in a second report by the pathologist Tagliaro on the amount of potassium in the vitreous fluid of the eye of the deceased are incorrect and in no way incriminating for DP.

{On July 7, 2017} the Court of Appeal of Bologna  acquitted her\footnote{\texttt{https://iscrivo.dcssrl.it/ISCRIVO/public/document/download?fileDoc=1D325AD31D471EC6EBA\\2FC758DA1816FB0DEE87FACE8299A006AD87992AD05AAE14889B8C0EA360C126023CB2649BF57\&public=true}}, overturning the first-degree sentence and attributed the death of RC to natural causes, declaring that ``the fact does not exist'' (in other words, a murder never took place). Daniela was then released from prison.  In 2018 the Supreme Court (Corte di Cassazione) annulled the sentence (``cassation'') and ordered the trial to be repeated. In 2019 upon appeal,  she was acquitted again, but in 2020 the Supreme Court ordered yet another new trial, an almost unprecedented situation in the Italian judicial system.

A further trial independently took place on the 15$^{\mbox {th}}$ of December 2020, at which Daniela Poggiali was sentenced to thirty years in prison, accused of murdering another one of her patients, this time 95-year-old Massimo Montanari (MM), the former employer of her  partner,  who died on March 12, 2014 (a month before the death of RC).

In two subsequent appeals in Bologna she was acquitted -- and at the end of the first one released after 1.003 days in prison - but with both acquittals overridden by Cassation.

In October 2021, DP was again on trial at the Court of Appeal of Bologna and Julia Mortera was called by the defence lawyers  to testify. Some parts of this paper are based on her testimony given on   October 24,  2021. The following day the president of the Court of Appeal of Bologna, Judge Stefano Valenti, acquitted Daniela yet again, now of both murder charges.

\section{Data}

\label{sec:data}
The Lugo hospital ward catered for terminally and seriously ill and elderly frail patients, so it was usual that on most days there were one or more deaths.  In these circumstances it is not unusual for nurses to appear to become quite indifferent to patients' deaths. 
In the case we are concerned with four particular wards or sectors (we will take the two words to be synonymous) of the ``Umberto I'' hospital,  A, B, C, and D.
Each consists of many small rooms each with only one or two beds, and they are all located on the same floor of the same building.
The building has two long wings, which branch off from a central part housing central facilities. A long corridor runs through each wing.
Sector A and B rooms were in one wing, first A and then B. Sector C and Sector D rooms were in the other wing, opposite to one another on each side of that wing's corridor (see Figure \ref{fig:str}). 

Most of the analyses reported here will be based on the \textit{baseline} period of two years of DP's employment from  9/4/2012 to 8/4/2014. Table \ref{tab:sector} shows the sectors where DP was on duty.
In 2012 DP was most frequently on duty in sector A, in 2013 in sector D, and in 3 months of 2014 in sectors A and C. DP is almost always assigned to a single sector and rarely to two contiguous sectors. In the TM analysis, deaths in both A and B and in both C and D are added and all the deaths in two contiguous sectors are allocated to DP, even when she is only working in one sector. The prosecution argument for this was that a malevolent nurse could easily slip over to a room in adjacent sector without anyone noticing. This might have been an opportunistic argument, found by trying out many different such groupings in order to find evidence which appears incriminating against DP, recall the \emph{Texas sharpshooter fallacy}.\footnote{
	https://en.wikipedia.org/wiki/Texas\_sharpshooter\_fallacy}  Whether or not this was the case, we do not know which statistical analyses were done by Micciolo before he wrote up the findings which were included in TM's report. It could be that he was effectively performing a forensic investigation, searching actively for evidence against DP for the public prosecutor, rather than also looking for evidence which could exonerate her (``tunnel vision'').

\begin{table}[ht]
	\centering
	\begin{tabular}{r|rrrr}
				& 2012 & 2013 & 2014 & total \\ 
		\hline
		A & 194 & 44 & 32 & 270 \\ 
		B & 3 & 4 & 1 & 8 \\ 
		AB & 0 & 1 & 0 & 1 \\ 
		C & 0 & 4 & 28 & 32 \\ 
		CD & 0 & 36 & 8 & 44 \\ 
		D & 0 & 169 & 2 & 171 \\ 
		\end{tabular}
	\caption{Distribution of sectors where DP was on duty.}
	\label{tab:sector}
\end{table}

Table \ref{tab:ricoveri1} shows the distribution of admissions per sector in  two comparable periods 9/4/2012  -- 8/4/2013  and 9/4/2013 -- 8/04/2014. The percentage of  deaths per admissions  $d/a$\% are  basically uniform over sectors and do not differ in the two periods. 

\begin{table}[htbp]
	\centering
	\begin{tabular}{l|cccc|cccc}
		period & \multicolumn{4}{c|}{9/4/2012  - 8/4/2013} & \multicolumn{4}{c}{9/4/2013 - 8/04/2014} \\
		\hline
		sector & A     & B     & C     & D     & A     & B     & C     & D \\
		\hline
		admissions ($a$) & 508   & 627   & 363   & 428   & 514   & 541   & 399   & 499 \\
		deaths ($d$) & 73    & 80    & 45    & 62    & 87    & 76    & 73    & 79 \\
		$d/a$ (\%) & 14\%  & 13\%  & 12\%  & 14\%  & 17\%  & 14\%  & 18\%  & 16\% \\
	\end{tabular}%
	\label{tab:ricoveri1}%
	\caption{Distribution of admissions and deaths per sector and year}
\end{table}%

Full time and fully trained nurses like DP worked on the following shifts: from 7:00 to 14:10; from 14:00 to 21:10, and from 21:00 to 7:10. Not infrequently, a nurse works on two consecutive shifts, for instance, afternoon and night. In practice (and as recorded for administrative purposes), the actual times nurses arrive and depart for their shifts is often different from the ``official'' times of the shifts. 

The ten minutes of overlap of the shifts was officially dedicated to the ``handover'', in which the departing nurses reported to the incoming nurses the condition of patients, information on new admissions, any deaths, other notable events. In practice, the overlap often took longer to complete.

\begin{figure}[htbp]
	\caption{Distribution of deceased patients' ages in Lugo hospital in all sectors, over the two year period 9/4/2012 -- 8/4/2014.}
	\begin{center}
		\includegraphics[width=1.1\textwidth]{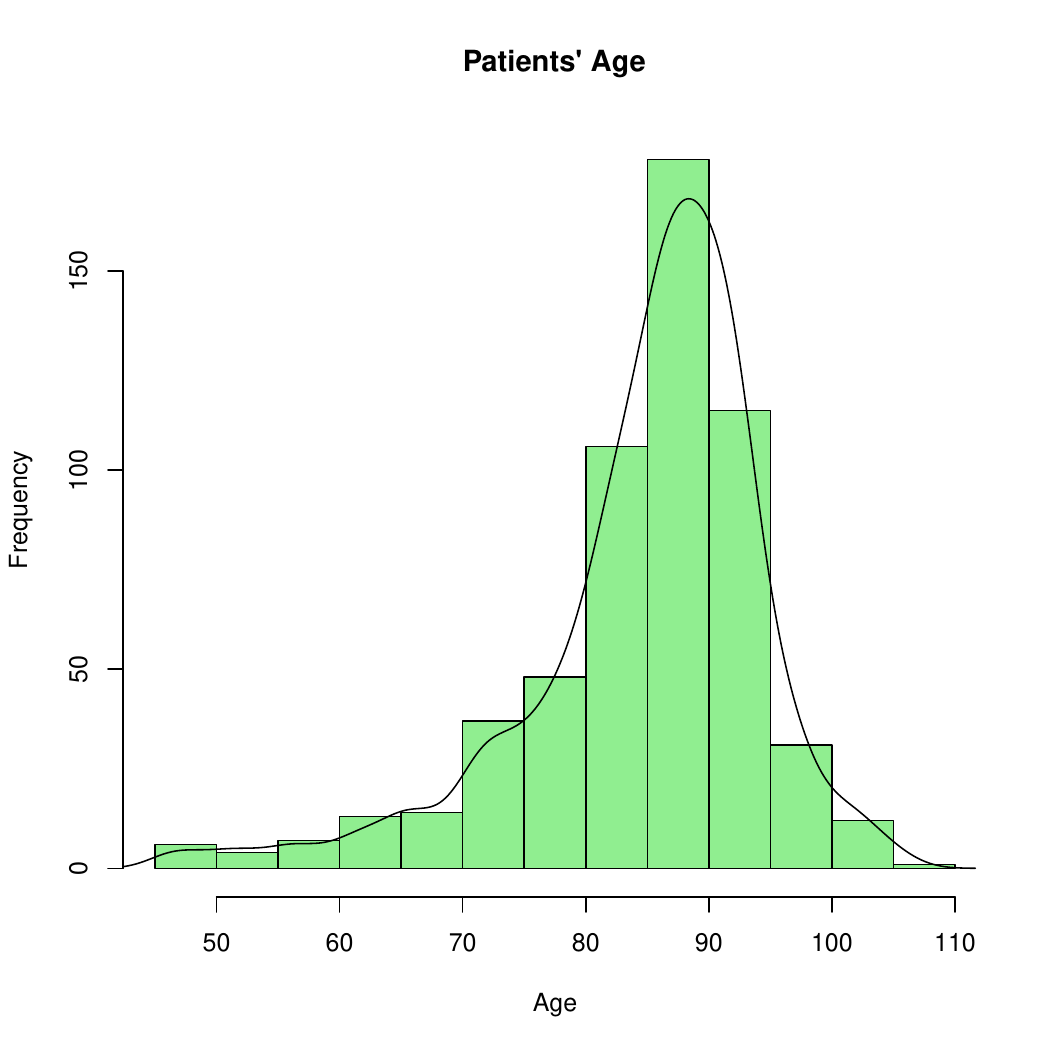}
		\label{fig:histage}
	\end{center}
\end{figure}

The patients in DP's wards were elderly or terminally ill. Figure \ref{fig:histage}  and Table \ref{tab:age}  show the age distribution of the deceased patients in all sectors during the period under examination. {The majority of the deceased patients are over 85 years of age with a median age of 87. }

\begin{table}[htbp]
	\centering
	{
\begin{tabular}{r|cccccc}
Age & (50,60] & (60,70] & (70,80] & (80,90] & (90,100] & $>100$ \\ 
\hline
Frequency &  11 &  27 &  85 & 283 & 144 &  13 \\ 
\end{tabular} 

\bigskip
\bigskip

\begin{tabular}{rrrrrrr}
	\hline
	& Min. & 1st Qu. & Median & Mean & 3rd Qu. & Max. \\ 
	\hline
	& 46.00 & 81.00 & 87.00 & 84.98 & 91.00 & 106 \\ 
	\hline
\end{tabular}
\caption{Distribution of deceased patients' ages in Lugo hospital in all sectors, over the two year period 9/4/2012 -- 8/4/2014.}
\label{tab:age}
}
\end{table}

\section{Analysis of mortality by time}
\label{sec:time}
\begin{figure}[htbp]
	\caption{Distributions of deaths per hour, minute, days and months between  9/4/2012--8/4/2014}
	\begin{center}
		\includegraphics[width=1.1\textwidth]{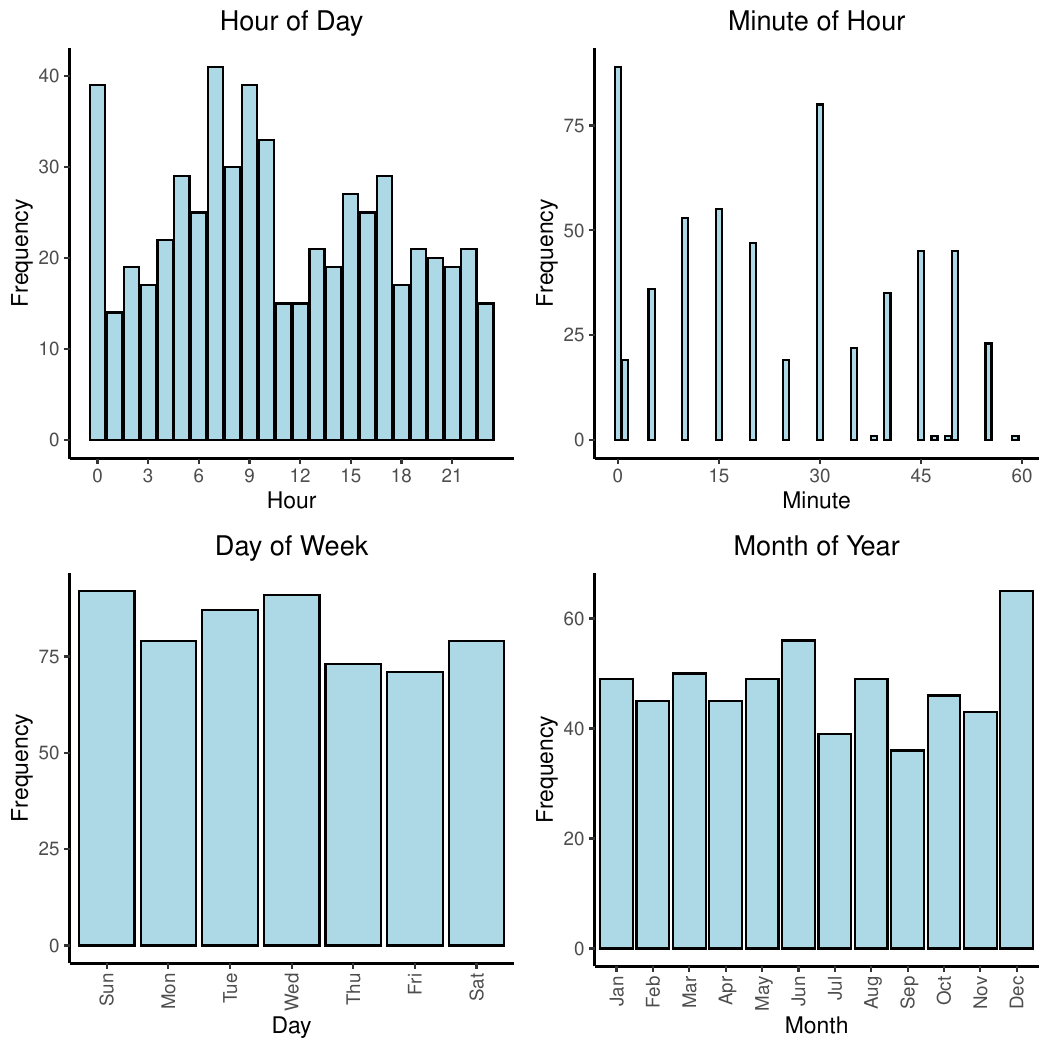}
		\label{fig:all}
	\end{center}
\end{figure}

\begin{figure}[htbp]
	\caption{Distribution of deaths per hour}
	\begin{center}
		\includegraphics[width=0.7\textwidth]{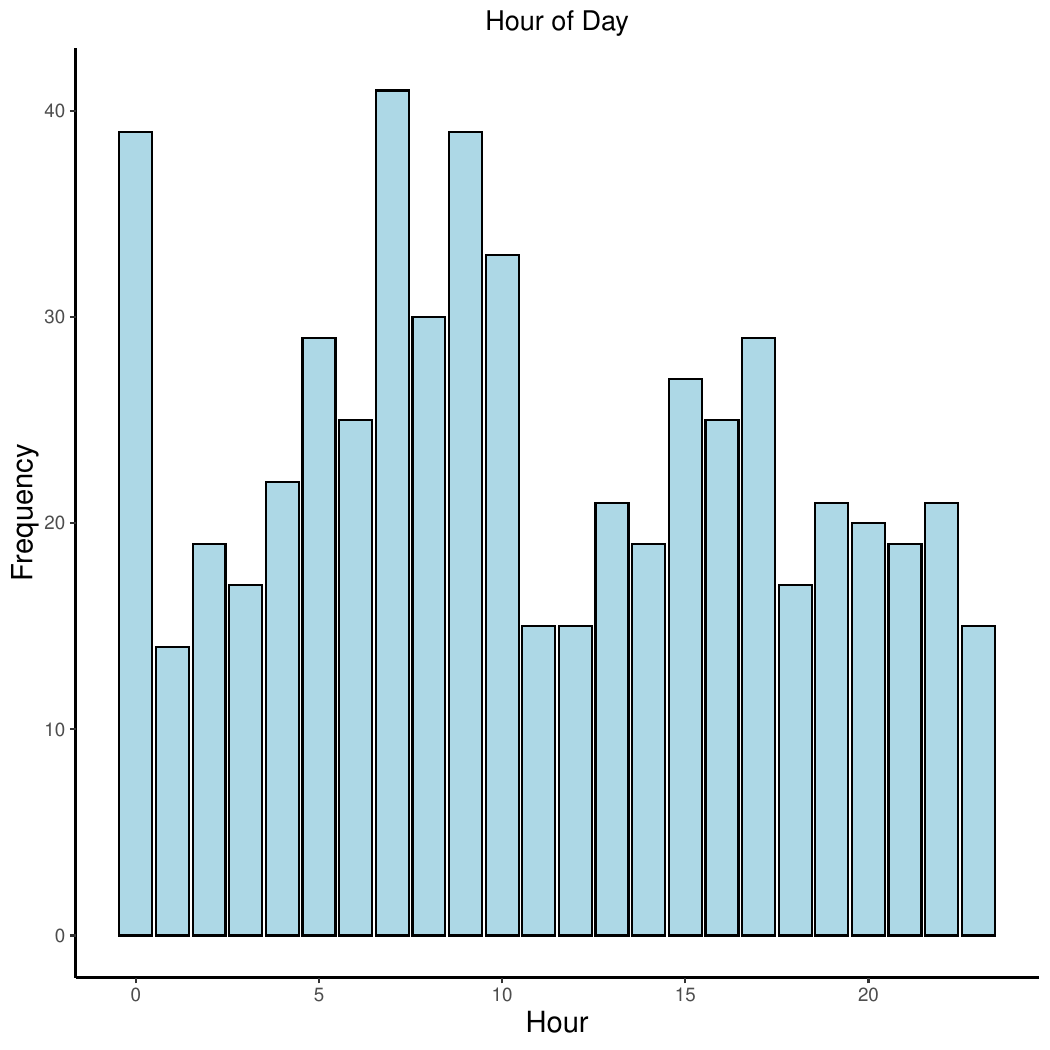}
		\label{fig:hours}
	\end{center}
\end{figure}

Figure \ref{fig:all} shows the distribution of the times of deaths recorded in different hours of the day, minutes of the hour, days of the week, and months of the year, over the baseline period of two years of DP's employment 9/4/2012--8/4/2014 (in the second half of this period the picture is the same). The top graph in the upper left corner of Figure \ref{fig:all}, and more clearly in Figure \ref{fig:hours}, shows that the hourly rate of deaths, or more accurately, of times a death is registered, is low during the evening and after midnight gradually increasing towards 7:00, but has two peaks at 24:00 (coded with 0) and at 7:00; it is highest in the morning after which it decreases again towards the evening.

The top right hand graph shows the distribution of the recorded minute within the hour of times of deaths. We can see that deaths are more likely to be recorded on the hour 
and at half past the hour. This indicates indeed that death times are not the actual death times but are rough times when a doctor certifies that death had occurred. This could be a confounding factor in calculating death rates. A further confounding factor can be the delay in registration of a death, whereby a death that occurred the day before is recorded as happening just after 24:00 (\ie, on the next day); see the peak found at midnight. 
There does not seem to be any difference in the day of the week when deaths are recorded (bottom left graph) whereas there is a peak of deaths in December (bottom right graph).

Confounding refers to phenomena which lead to biases in \emph{comparison} of different nurses. The question is whether or not these errors might lead to some nurses looking worse than others. The prosecution might claim that the errors do not affect the comparison, since they are the same for all nurses. But are they?

This could be a further inaccuracy inherent in how TM calculated the \emph{daily} mortality rate of each nurse.
Nurses on the night shift (21:00-24:00 \emph{and} 0:01-7:00) are calculated as  being on duty on two consecutive ``days'' in TM's calculations. Daily deaths might be assigned to a given nurse, even if they occurred during time slots in which she was not present. 

\begin{figure}[htbp]
	\centering
	\includegraphics[width=.7\textwidth]{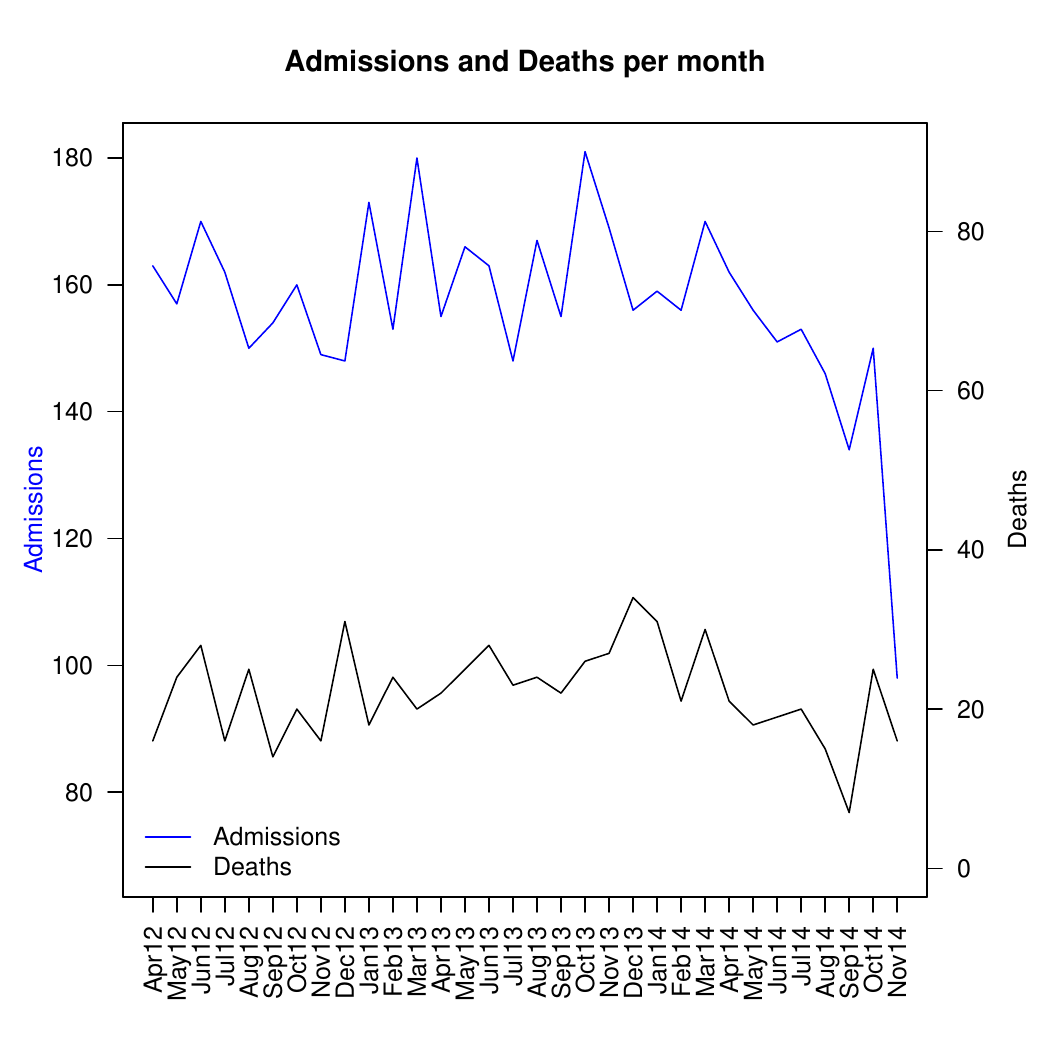}
	\caption{Distribution of the number of admissions and the number of deaths between  9/4/12 and 30/11/14.}
	\label{fig:ricxdec}
\end{figure}

Figure \ref{fig:ricxdec} shows the monthly distribution of admissions and deaths for the period 9/4/12--30/11/14. Note that after April 2014 the admissions dropped considerably (Lugo hospital had become infamous due to media coverage) and consequently deaths diminished. TM never took into account how many patients were admitted to the wards at any given time and took the decrease in deaths after DP was not working in the hospital as evidence against her. Furthermore, there is a spike in admissions in October 2013 followed by a spike in deaths in the following months.  Here one can see another spike in March 2014 with a corresponding spike in deaths in the same month. It is exactly in the latter period that TM note a large excess in the mortality rate for DP.    

\begin{figure}[htbp]
	\centering
	\includegraphics[scale=.9]{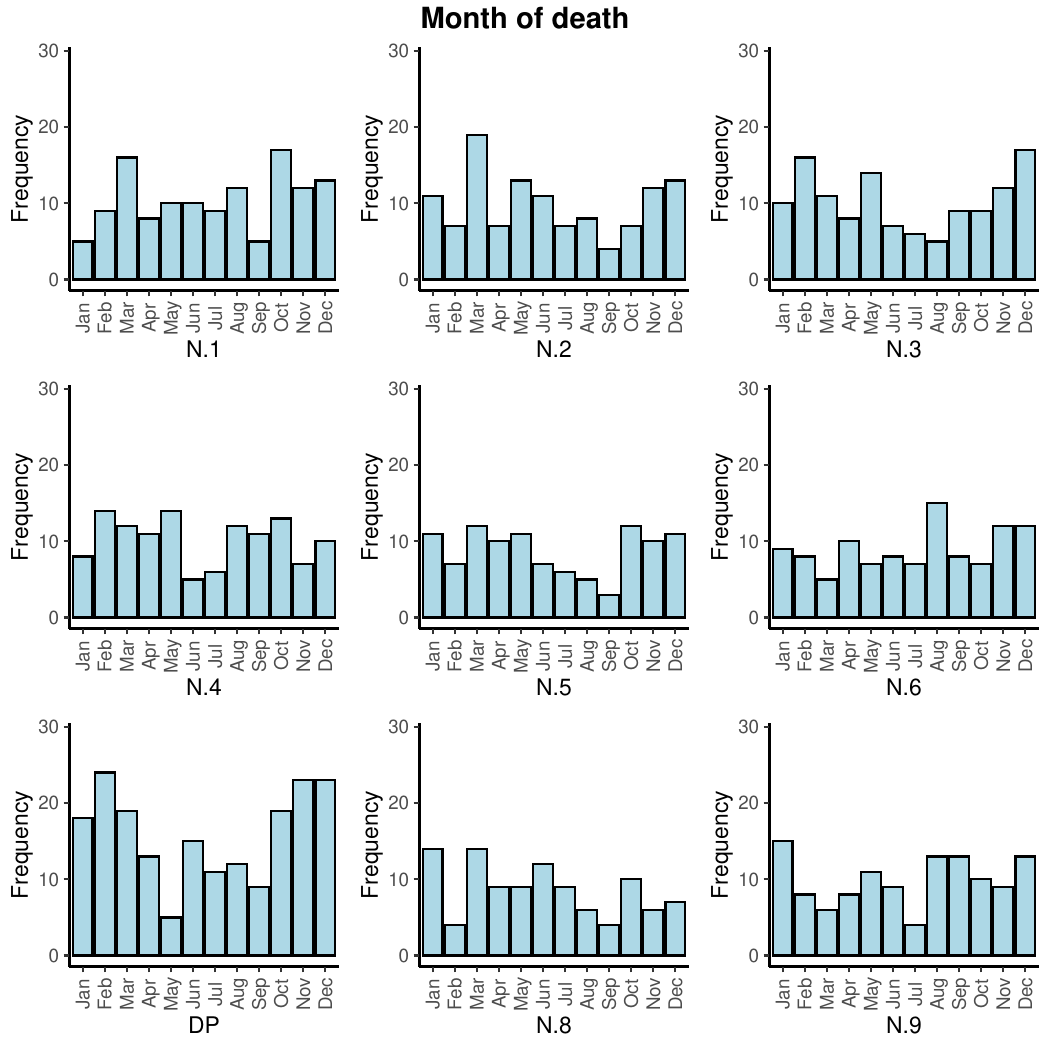}
	\caption{Distribution of deaths by month for those nurses who work a similar number of hours to DP.}
	\label{fig:mesixinf}
\end{figure}
\begin{figure}[htbp]
	\centering
	\includegraphics[scale=.9]{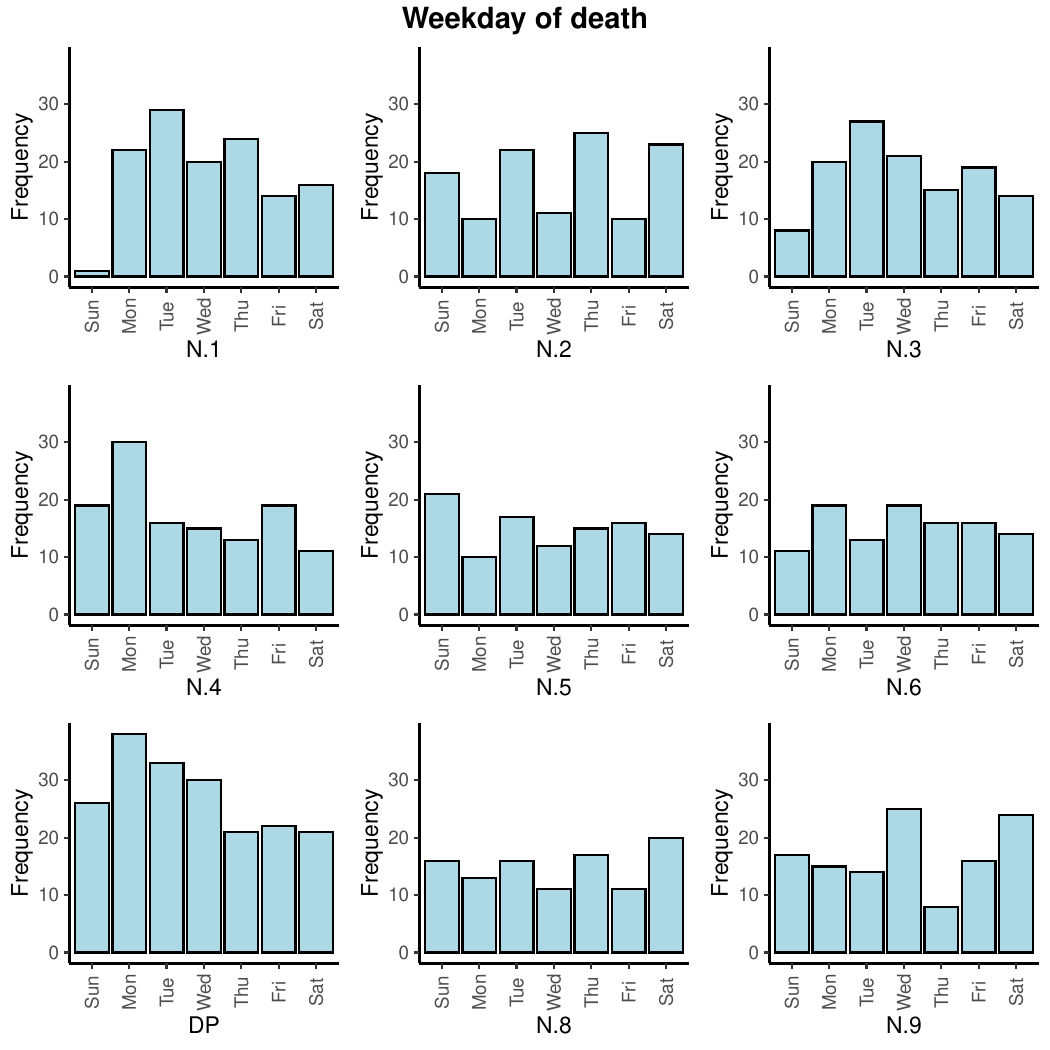}
	\caption{Distribution of deaths by day of the week for those nurses who work a similar number of hours to DP.}
	\label{fig:giornixinf}
\end{figure}

\begin{figure}[htbp]
	\centering
	\includegraphics[scale=.9]{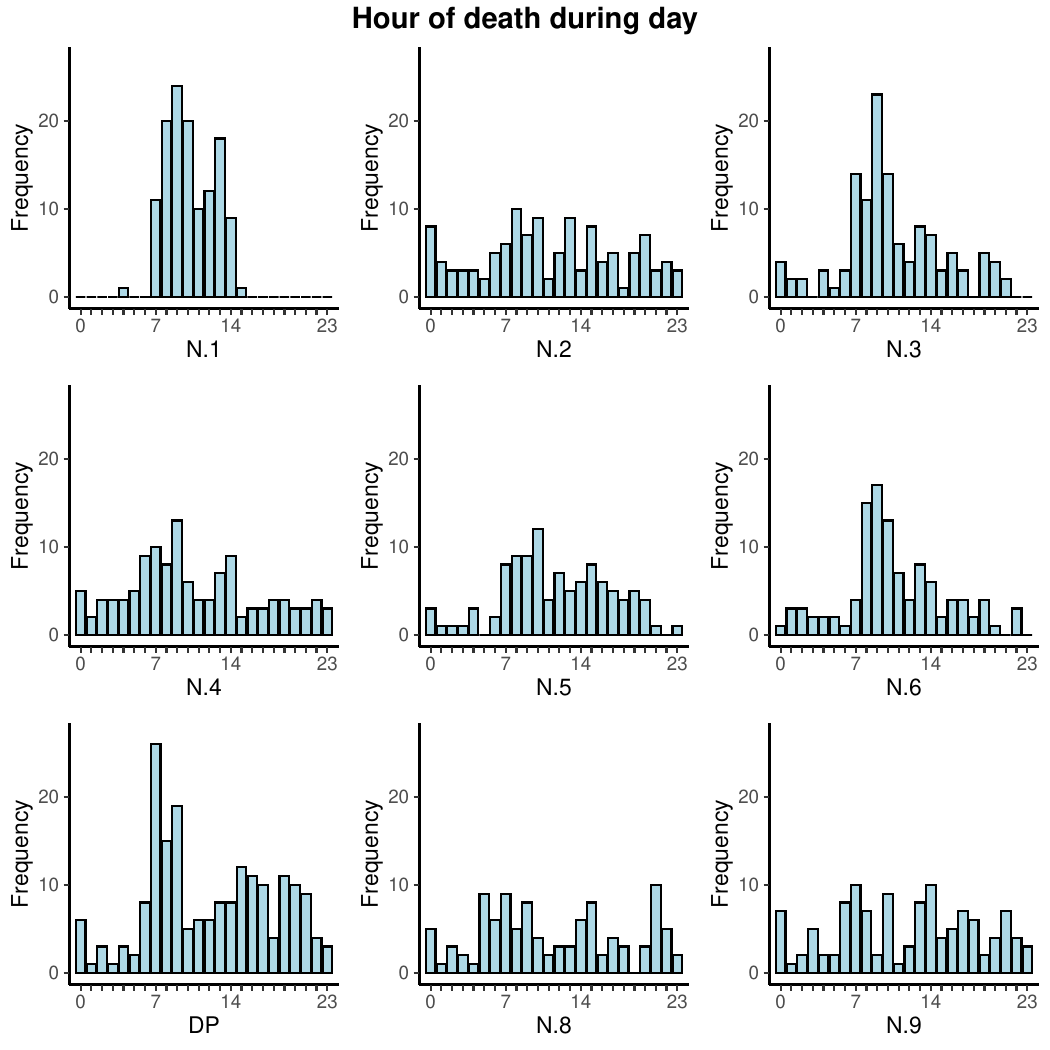}
	\caption{Distribution of deaths by hour for those nurses who work a similar number of hours to DP.}
	\label{fig:oraxinf}
\end{figure}

Figures \ref{fig:mesixinf}, \ref{fig:giornixinf}, and \ref{fig:oraxinf} show the distributions by month, day of week, and hour of death when a nurse is on duty, for only those nurses with a similar number of hours on duty as DP. Note that in Figure \ref{fig:mesixinf} (bottom left) DP has a high rate of deaths in the winter months precisely when deaths in general are highest. Figure \ref{fig:oraxinf} (bottom left) shows that DP, compared to the other nurses, has a peak of deaths between 6 and 7 a.m., precisely when the peak mortality rate is recorded (as shown in Figure \ref{fig:hours}).  From these plots we can see that also other nurses  have some unusual patterns of recorded deaths, but these should not be cause of suspicion. 

The number of deaths varies between months and across years (see Figure \ref{fig:ricxdec}) and between different sections of the hospital (see Table \ref{tab:ricoveri}). There is an increase in the number of deaths in  C in the second year. Sections A and D have 17 beds, B has 18 beds and C is slightly smaller with 14 beds. We do not know whether more critically ill patients are assigned to  C.
This may explain why TM find such an \emph{unfavourable} outcome for this nurse. 

TM attempted to take account of confounding factors due to daily, seasonal or longer term trends, by always comparing the death rate experienced by DP on days that she worked in the pair of sectors where she was working, against the pair of sectors where she was not working. This assumes that the two pairs of sectors are exactly comparable in terms of numbers of patients and severity of their illness. Moreover, it neglects the fact that during a full three shift cycle, a full-time nurse will tend to be found when and where most nurses are working, namely in the morning and in the area with most nurses on duty. In further analyses they compared DP with nurses who are as similar as possible in terms of number of hours worked. Even then however, confounding factors still remain which are not taken into account in their analyses.

\section{Shifts}
\label{sec:shifts}

\begin{figure}[htbp]
	\centering
	\includegraphics[scale=.75]{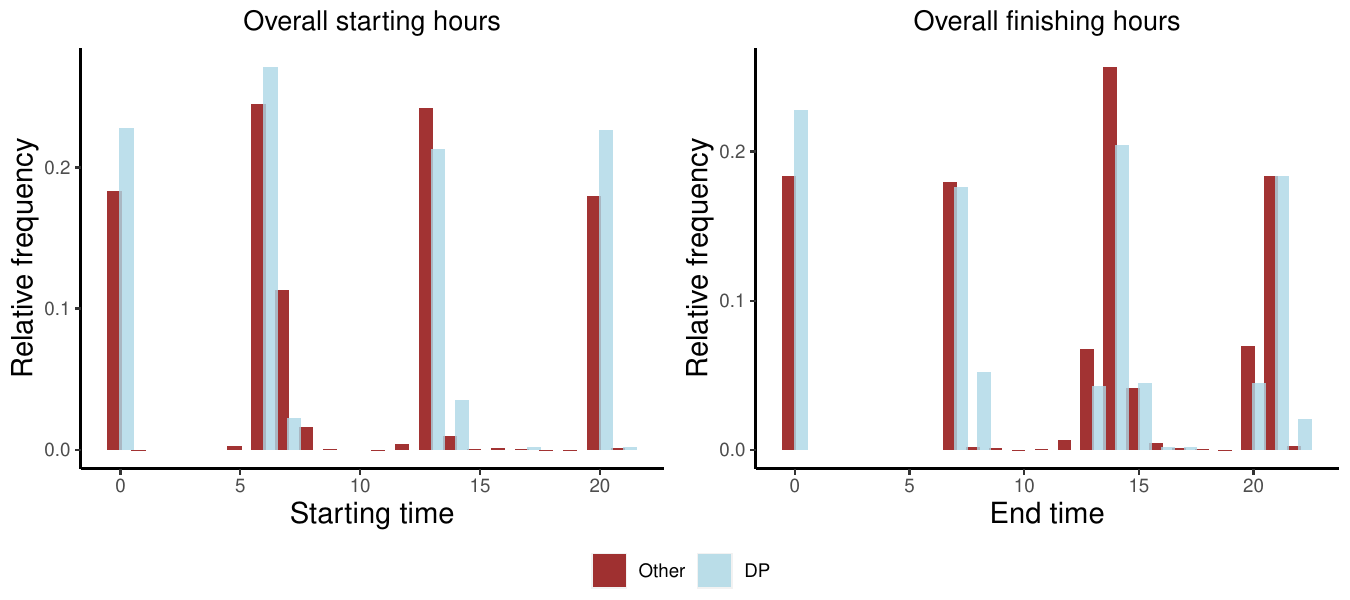}
	\caption{(left) Distribution of starting times for nurses (blue)  and those of DP (red), (right) the same for finishing times.}	
\label{fig:startend}

\end{figure}

The histogram on the left in Figure \ref{fig:startend} shows the distribution of nurses' shifts' starting times (coloured in red) versus those of DP (coloured in light blue), whereas the one on the right shows the distribution of nurses' shifts' finishing times (coloured in red) versus those of DP (coloured in light blue).

Table \ref{tab:iniz} shows the distribution of starting time for DP. She starts the shift more often at 6:00 and 24:00 (coded as 0) so she is associated with the peak in deaths between 7:00 and 7:05 and between 24:00 and 00:05 (see the top two graphs in Figure \ref{fig:all}). Recall that   shifts are from 7:00 to 14:10; from 14:00 to 21:10, and from 21:00 to 7:10. 

Notice that she starts 123 times at midnight. This is merely the continuation of a shift or even a double shift which started earlier -- in the evening at 20:00 or 21:00 (123 shifts) or even in the afternoon at 13:00 or 14:00 (134). Notice also that when DP, compared to the other nurses, starts her shift in the morning, she arrives very early at 6:00 when she has the maximum number of shifts (146) and when she ends her shift in the evening she stays well past the end of her shift. Recall that there is a spike in the number of deaths in the morning at 7:00 and at 24:00 hours, as shown in Figure \ref{fig:hours}. 

\begin{table}[ht]
	\centering
	\begin{tabular}{r|llllllll}
		start of shift & 0 & 6 & 7 & 13 & 14 & 17 & 20 & 21 \\ 
		\hline
		frequency & 123 & 146 & 12 & 115 & 19 & 1 & 122 & 1 \\ 
			\end{tabular}
\caption{Distribution of starting times for DP's shifts.}
\label{tab:iniz}
\end{table}

\begin{table}[ht]
	\centering
	\begin{tabular}{r|rrc}
		
	sector	& \multicolumn{1}{c}{admissions} & \multicolumn{1}{c}{deaths}  & \multicolumn{1}{c}{deaths/admissions (\%)} \\ 
		\hline
		A & 1025 & 159 & 16\% \\ 
		B & 1168 & 154 & 13\% \\ 
		C & 762 & 118 & 15\% \\ 
		D & 930 & 141 & 15\% \\ 
		\hline
	\end{tabular}
\caption{Distribution of admissions and deaths per sector.}
\label{tab:ricoveri}
\end{table}

From the Table \ref{tab:ricoveri} we see that sectors A + B have a higher number (about 25\%) of admissions than sectors C + D. On the other hand, the fraction of deaths is similar for all sectors varying between 13\% and 15\%. 
DP worked most often in sector A, followed by sector D.

\section{Number of nurses on duty and corresponding number of deaths}
\label{sec:service}
As in all hospitals, nurses do not work alone, but there are several nurses on duty at the same time,  together with medical, auxiliary, and other health care workers.
\begin{table}[htb]
	\centering
\begin{tabular}{r|rrrrrrrrrrrr}
number of nurses	& 3 & 4 & 5 & 6 & 7 & 8 & 9 & 10 & 11 & 12 & 13 & 14 \\ 
	\hline
number of deaths &   1 & 191 &  64 & 114 &  27 &  51 &  82 &  21 &  10 &   4 &   6 &   1 \\ 
	\end{tabular}
	\caption{Distribution of the number of nurses on duty and corresponding  number of deaths.}
	\label{tab:infPres}
\end{table}

\begin{table}[htb]
	\centering
	\begin{tabular}{r|rrrrrrrrrrrr}
	number of nurses & 3 & 4 & 5 & 6 & 7 & 8 & 9 & 10 & 11 & 12 & 13 & 14 \\ 
	\hline
	number of deaths &   1 &   8 &  57 & 113 &  25 &  48 &  79 &  19 &  10 &   4 &   6 &   1 \\ 
	\end{tabular}
	\caption{Distribution of the number of nurses on day shifts and corresponding  number of deaths.}
	\label{tab:infPresG}
\end{table}

\begin{table}[htb]
	\centering
\begin{tabular}{r|rrrrrrr}
	
number of nurses	& 4 & 5 & 6 & 7 & 8 & 9 & 10 \\ 
	\hline
	number of deaths & 183 &   7 &   1 &   2 &   3 &   3 &   2 \\ 
	
\end{tabular}
	\caption{Distribution of the number of nurses on night shifts and corresponding  number of deaths.}
	\label{tab:infPresN}
\end{table}

Table \ref{tab:infPres} shows the distribution of the number of nurses present and the relative number of deaths and shows that there are many nurses present at the same time. The data provided by the Local Health Unit (AUSL) on the time-stamping of the nurses' shifts (they have to clock in when they arrive for work, and to clock out when they leave). 
 There is an  average (and median) number of nurses  on duty of  around   6  (with a standard deviation around 2) and with a range that varies from a minimum of 3 to a maximum of 14.

Tables \ref{tab:infPresG} and \ref{tab:infPresN}
 show the distribution of the number of nurses present and relative number of day and night time deaths. Note that on night shifts the number of nurses present is less than on day shifts.

\begin{table}[ht]
	\centering
	
\begin{tabular}{r|rrrrrrrrrrrr}
	
number of nurses	& 3 & 4 & 5 & 6 & 7 & 8 & 9 & 10 & 11 & 12 & 13 & 14 \\ 
	\hline
	n. deaths others &   1 & 163 &  44 &  61 &  12 &  32 &  43 &  16 &   5 &   2 &   2 &   0 \\
		\hline 
	n. deaths DP &   0 &  28 &  20 &  53 &  15 &  19 &  39 &   5 &   5 &   2 &   4 &   1 \\ 

\end{tabular}
	\caption{Distribution of the number of nurses on duty with and without DP and corresponding number of deaths.}
	\label{tab:altreDP}
\end{table}
\begin{figure}[htbp]
	\centering
	\includegraphics[width=.7\textwidth]{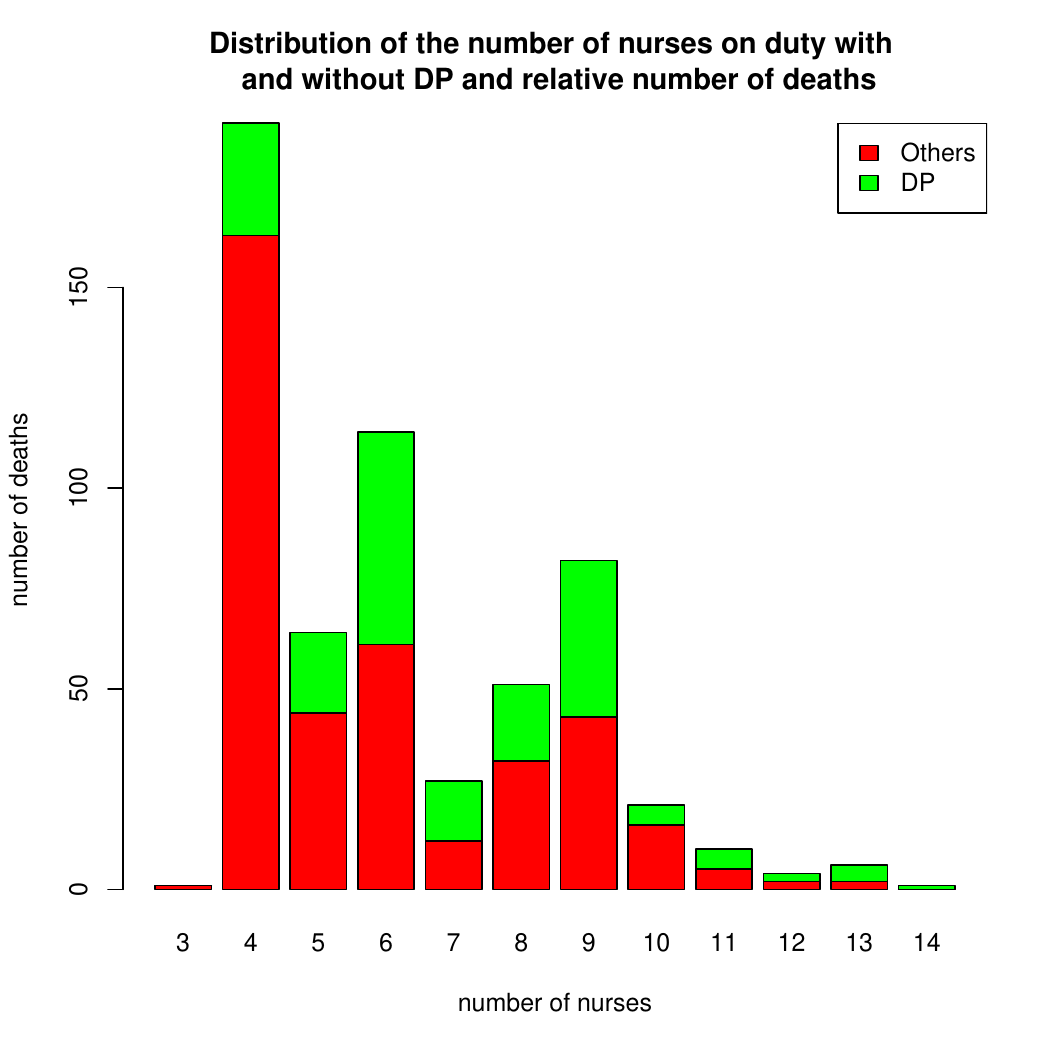}. 
	\caption{Distribution of the number of nurses on duty with and without DP and corresponding number of deaths.}
	\label{fig:infeDP}
\end{figure}
\begin{table}[ht]
	\centering
\begin{tabular}{r|rrrrrrrrrrrr}
	
number of nurses	& 3 & 4 & 5 & 6 & 7 & 8 & 9 & 10 & 11 & 12 & 13 & 14 \\ 
	\hline
	n. deaths others &   1 &   3 &  40 &  61 &  10 &  31 &  42 &  14 &   5 &   2 &   2 &   0 \\ 
	\hline
	n. deaths DP &   0 &   5 &  17 &  52 &  15 &  17 &  37 &   5 &   5 &   2 &   4 &   1 \\ 
	
\end{tabular}
	\caption{ Distribution of the number of nurses on  duty during the day with and without DP and corresponding number of deaths.}
	\label{tab:altreDPD}
\end{table}
\begin{table}[ht]
	\centering
\begin{tabular}{r|rrrrrrr}
	
number of nurses	& 4 & 5 & 6 & 7 & 8 & 9 & 10 \\ 
	\hline
		n. deaths others & 115 &   4 &   1 &   2 &   1 &   3 &   1 \\ 
	\hline
	n. deaths DP &  68 &   3 &   0 &   0 &   2 &   0 &   1 \\ 
\end{tabular}
	\caption{Distribution of the number of nurses on duty during the night  with and without DP and corresponding number of deaths.}
	\label{tab:altreDPN}
\end{table}
Figure \ref{fig:infeDP} and Table \ref{tab:altreDP} show the distribution of the number of nurses on duty  with and without DP and the corresponding numbers of deaths. It can be clearly seen that when deaths occur with DP present, 6, 7, and 8 nurses are most often on duty (in all four sectors combined), and rarely only 3. This fact, in our opinion, confirms that, as there are often times many nurses on duty when a death is recorded, a single nurse cannot be considered solely responsible for a given death.

\begin{figure}[htbp]
	\centering
	\includegraphics[width=.7\textwidth]{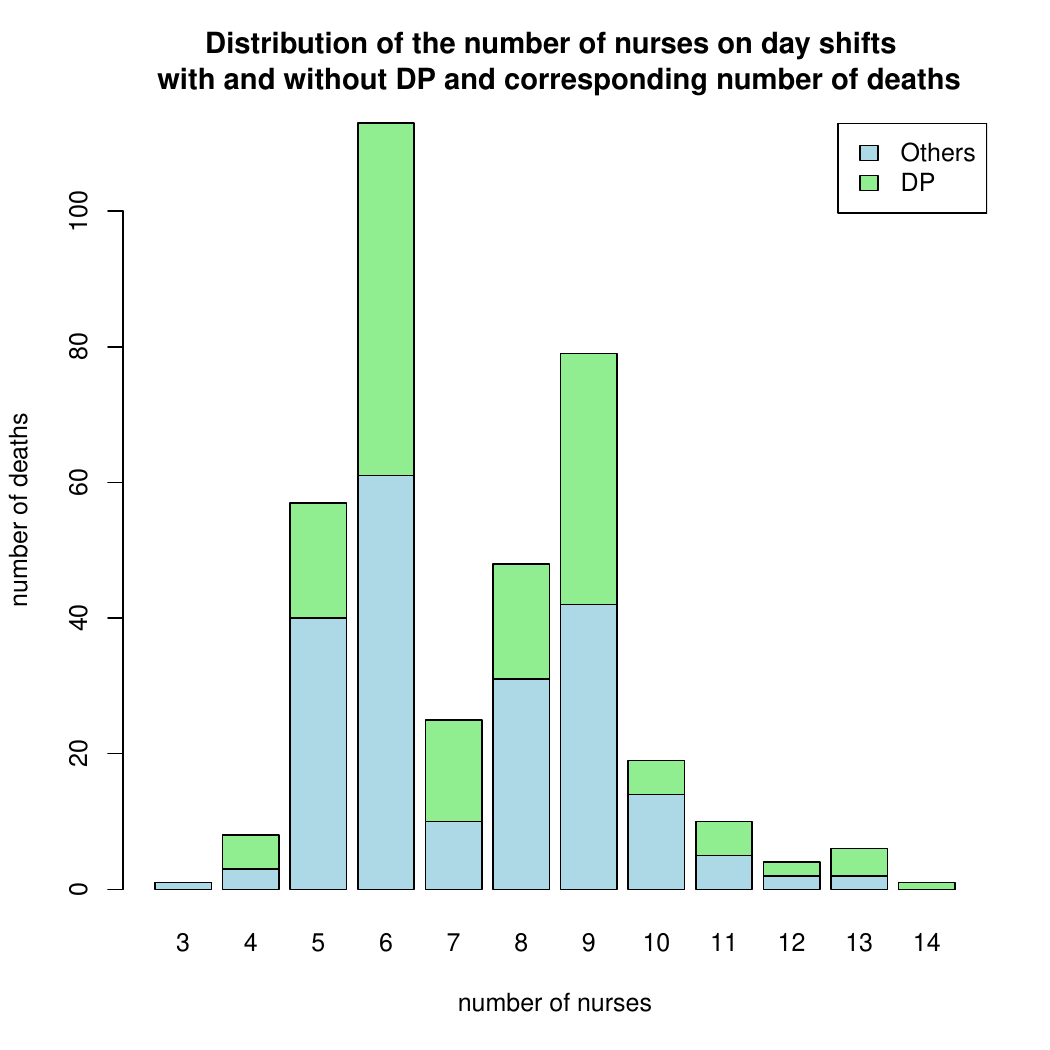}
	\caption{Distribution of the number of nurses on day shifts  with and without DP and corresponding number of deaths.}
	\label{fig:infeDPG}
\end{figure}

\begin{figure}[htbp]
	\centering
	\includegraphics[width=.7\textwidth]{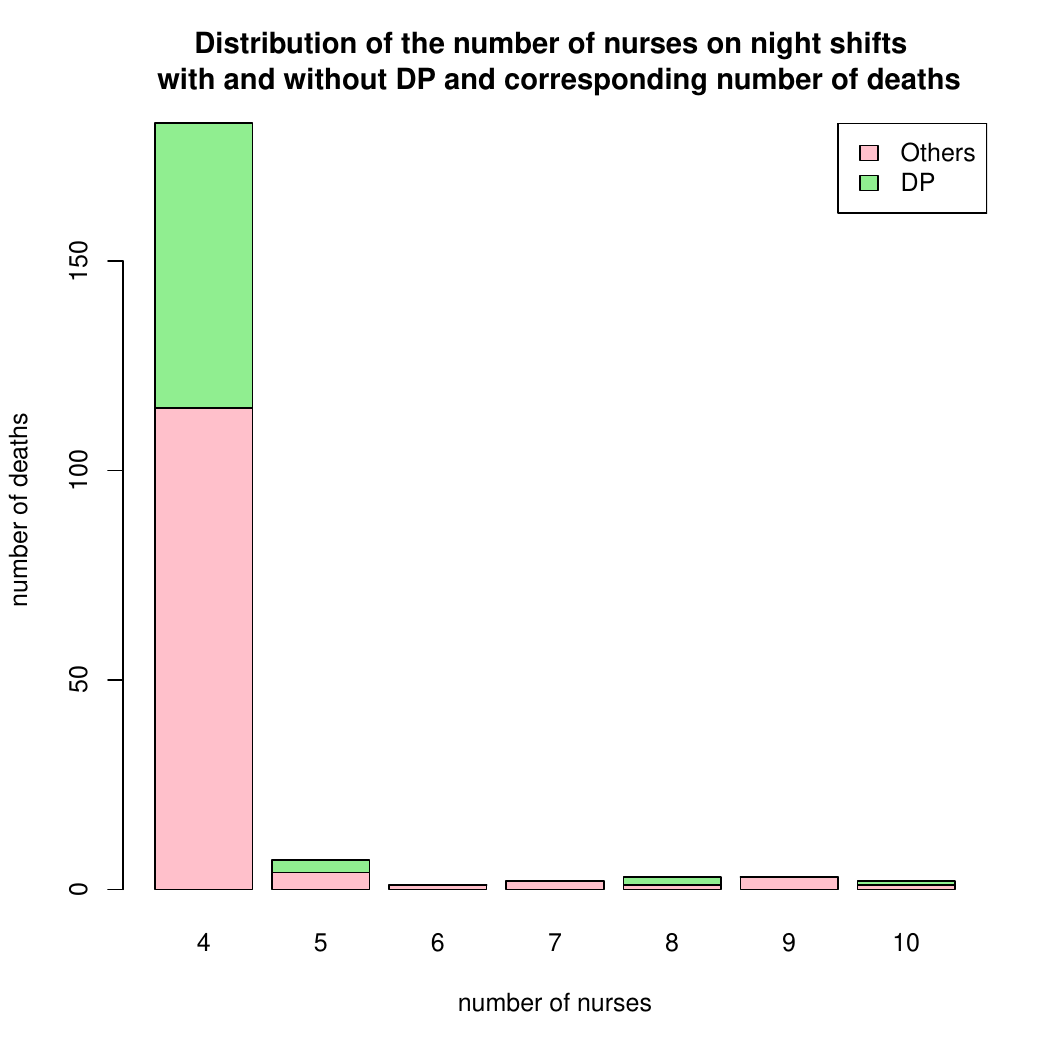}
	\caption{Distribution of the number of nurses on night shifts  with and without DP and corresponding number of deaths.}
	\label{fig:infeDPN}
\end{figure}

Tables \ref{tab:altreDPD} and \ref{tab:altreDPN} and Figures \ref{fig:infeDPG} and \ref{fig:infeDPN} illustrate how many nurses are present with DP in \textit{day shifts} and \textit{night shifts} when deaths occur. Recall that it is in the day shifts that the greatest number of deaths occur. Comparing Figure \ref{fig:infeDPG} with Figure \ref{fig:infeDPN} shows that in day shifts compared to night shifts there are many more nurses on duty and this is precisely when many more deaths are observed and when the DP mortality rate is significantly high.

\section{Bayes's rule}
\label{sec:bayes}

TM write (our translation): \textit{Quantification takes place in the form of probability, through the calculation of the so-called $p$-value. Small values of the $p$-value (i.e., small probability values) exclude a pure chance effect, while values that are not small do not allow one to exclude a pure random effect.}  [~\dots~]  \textit{the level of significance  represents a guarantee against the risk of drawing a wrong conclusion that a non random difference exists (and therefore there is a causal effect). The smaller the level of significance, the higher the level of assurance.}

This explanation may lead the reader to interpret the $p$-value as the probability of the null hypothesis (the null hypothesis is the defence hypothesis). The $p$-value is the probability of finding the observed value or an even more extreme value. It is not the probability that the observed (high) death rate is due to random fluctuations. TM seem to interpret a strong association as implying a direct causation. This is a very serious and dangerous interpretation as it can lead to condemning an innocent suspect. Correlation or association between two variables generally has a causal explanation, but the causation might be in either direction, or it might be due to a common cause in the past, or it might be due to selection in the future according to a common consequence of both variables.

To explain the previous statements, assume that an unexpectedly large number of deaths occur among patients when a particular health care professional is on duty.
Suppose further that an expert concludes that the probability that so many deaths occur by chance is only 0.000001, or one in 1 million. What can we conclude about the probability that the health care professional committed a crime?
There are a number of potential problems that arise in calculating $p$-value
\citep{benjamin2019three,berger1991interpreting}. 
What conclusions can we draw in this case? It is often mistakenly believed that a low $p$-value indicates a high probability of the hypothesis that misconduct has happened. One might jump to the conclusion that this $p$-value means that there is only one possibility in a million that so many deaths happened by chance, and correspondingly almost certain that the deaths are due to some other cause. If medical misconduct seems be the only plausible alternative explanation, then one might be tempted to conclude that the probability of healthcare misconduct is overwhelming (999999 chances in 1 million). In this way, a $p$-value of 1 in 1 million can be mistakenly taken as evidence that there is only a 1 in 1 million chance that the health care provider is innocent. This error of interpretation seems to be ubiquitous in the reports of TM. 

When assessing the probative value of an item of evidence Bayes' rule derives what can be logically inferred from it. In a criminal proceeding two hypotheses can be identified:
\begin{description}
	\item[$H_d$] the defence hypothesis (or null hypothesis) that no misconduct has taken place;
	\item[$H_p$] the prosecution hypothesis (or alternative hypothesis) that the health care professional has engaged in misconduct that places
	his/her patients at a high risk of death.
\end{description}

Let $E$ be the evidence for a specific number of patient deaths in a given period. What
investigators/judges/jurors want to know is the probability that $H_p$
is true in light of the evidence, \ie, the posterior probability  $P(H_p|E)$. The $p$-value
provided by TM is not the value of $P(H_p|E)$. Rather, it is the probability of the observed (or even more extreme) evidence, under the presumed hypothesis of the defence, $H_d$. Research \citep{thompson1987interpretation} has shown that people frequently fall victim to the logical fallacy known as transposition of the conditional, \ie, they confuse or equate the probability of the evidence given a hypothesis with
the probability of a hypothesis given the evidence ($P(E|H)$ with $P(H|E)$). In judicial terms this  is called the ``prosecutor's fallacy''  because it typically produces seemingly convincing evidence of guilt. The prosecutor's fallacy is a seductive and widespread mode of reasoning, affecting the general public, the media, lawyers, jurors
and judges alike. 
Cases in which the prosecution fallacy may have contributed to judicial errors that led to the conviction of an innocent person
are, among others, the case of Sally Clarke (Royal Statistical Society (23 October 2001) ``Royal Statistical Society concerned by issues raised in
Sally Clark case'')  and that of Lucia De Berk \citep{meester2006ab}.

In principle, a judge might try to assess his or her posterior probabilities of the prosecution and defence hypotheses  $H_p$ and $H_d$ conditional on evidence $E$: \ie, $P(H_p|E)$ and $P(H_d|E)$. These posterior probabilities can be
constructed out of other, more basic, ingredients, specifically:
the probabilities of the evidence given the hypotheses, namely $P(E|H_p)$ and  $P(E|H_d)$; and the prior probabilities $P(H_p)$ and  $P(H_d)$, \ie, the probabilities of  $H_p$ and  $H_d$ before any evidence is incorporated.
Bayes's rule -- a trivial consequence of  the definition of conditional probability -- 
tells us that the a posteriori odds are given by the prior odds times the ratio of the probabilities of the evidence given each of the hypotheses (the likelihood ratio):
\begin{equation}
	\label{eq:bayes}
	\frac{P(H_p  |  E)}{P(H_d  |  E)} = \frac{P(H_p)}{P(H_d)}
	\times \frac{P(E  |  H_p)}{P(E   |  H_d)}.
\end{equation}
Despite the unquestionable correctness of Bayes' rule,
it is often replaced by other probabilistic and non-probabilistic arguments, which can be very misleading and dangerous. There is of course an immense legal literature on whether or not Bayesian reasoning may be used in criminal cases, see for instance \url{https://academic.oup.com/lpr/article/5/2/167/927735}. Some legal scholars argue that probabilistic reasoning is not admissible for a variety of reasons, for instance, jurors can never understand it, or because of other matters of principle.

In the case under consideration an essential factor, in addition to the probabilities of the evidence given the
hypotheses, is  the a priori (or initial) probabilities of the hypotheses. 
In order to  consider the possibility that DP is responsible for the high rate of
mortality in Lugo hospital,  one might assess how many cases exist in Italy of nurses
who deliberately killed patients. In Italy, there has been only one case in 50 years
of a nurse convicted of killing 5 or more patients. 
In that case there was concrete evidence of guilt: she was, in fact, caught red-handed and confessed. (In the whole world in 100 years there has been only one nurse who has been charged with killing
50 or more patients. That was Niels H\"ogel, who was caught red handed (in \textit{flagrante delicto} and confessed).

The number of nurses in Italy in 2014 was 280,000 (source: the Italian Statistical Institute, Istat). Based on this,
we can estimate the possibility of a nurse in Italy killing patients in one or two
years of service and we arrive at a probability of about 1 in 14 million, that is 0.000007\%.  Researchers from criminology and forensic psychology, estimate a prior probability at around 1 in two million as a good guess for the whole developed world. It can only be a very rough guess: on the one hand, some convictions are disputed, on the other hand, there might be cases which have not been noticed at all\footnote{https://dokumen.tips/documents/diss-mas-3.html}.

\begin{figure}[htbp]
	\centering
	\includegraphics[width=.8\textwidth]{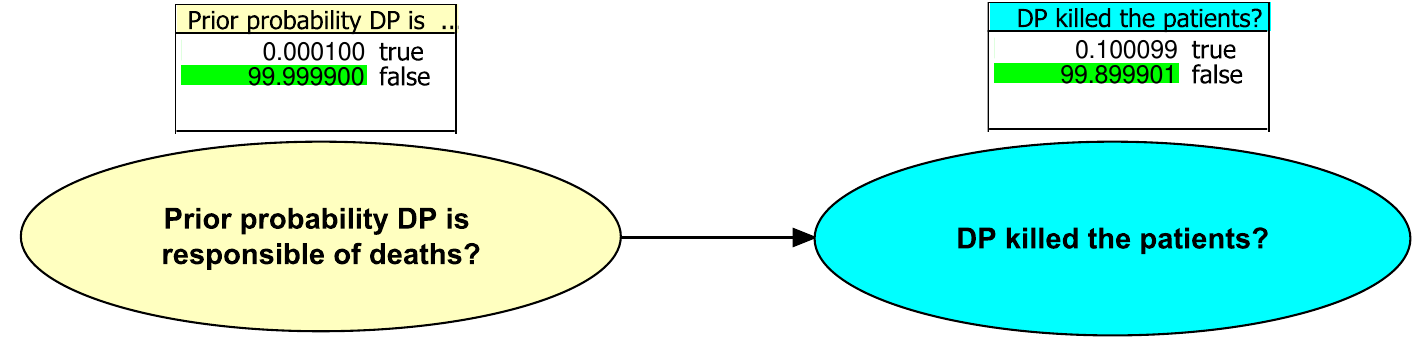}
	\caption{Representation of Bayes's rule starting from the a priori probabilities of the hypotheses (yellow node) we arrive at the a posteriori probabilities of the hypotheses (red node), given evidence $E$. The values of the probabilities are given in percentages}.
	\label{fig:BN}
\end{figure}

Figure \ref{fig:BN} illustrates the logical and consistent reasoning based on Bayes's rule. Assuming a priori probability $H_p$ of 1 in a million, \textit{i.e}.  0.0001\% --
which is much less favourable than the estimated value of 0.00007\% -- the a posteriori probability of
guilt is $P(H_p|E) =0.001=0.1\%$. This corresponds to one case out of 1000, certainly not beyond any reasonable doubt.
In the calculation we have used the evidence, in line with the results of TM, given by
$P(E|H_p)$  = 0.99,  and  $P(E|H_d)$ = 0.001 or 0.1\%.

Even assuming a very high a priori probability $P(H_p)=0.02= 2\%$ (this implies that 2\% of all nurses in Italy are ``angels of death''), totally unfavourable to DP, we have, on the basis of the evidence reported by TM, a slightly higher a posteriori probability that DP is guilty, equal to 2.08\%.  This illustrates that even using such high evidence on the basis of mortality rates as TM indicate, the final probability differs very little from the a priori probability, indicating that the conclusions reached by TM are totally misleading.

Moreover, TM seem to imply that finding an association necessarily implies causation. They explicitly state (our translation): \textit{In a certain sense the \textit{significance level} represents a ``guarantee'' against the risk of drawing a wrong conclusion, claiming that there is a non-random difference (and thus a cause), when in fact this difference is still of random origin. The smaller the level of significance, the higher the level of ``guarantee'' that is sought}. Yet as we have said, an association between two variables could be due to other factors that
influence both variables, or even due to selection based on variables which are common consequences.
In the report Tagliaro and Micciolo (TM) talk about ``association'' without clarifying that the term \emph{cannot be interpreted as synonymous with causality}. A reader, in our opinion, could be led to interpret the term as synonymous with causality. In general, there is no correlation without causality, but the causality might be due to common causes or common consequences, both known and unknown.

Preliminary remarks by the medical director of the AUSL (the local health administration), Dr G. Spagnoli, when comparing  the mortality rate in the medical ward of the hospital of Lugo with that of  similar wards in Faenza and Ravenna (managed by the same AUSL) claimed that there were no significant differences (see page 6 of the minutes of the hearing 108521 of 23/10/2015 Court of Ravenna).

To analyse the data properly, it would be necessary to construct a complex model that takes into account not only individually all of the measured and measurable confounding variables (hours of attendance and not on duty, deaths on duty and not on duty, number of admissions, hours and shifts of attendance, area of service, number of other nurses present when a death occurs, severity of deceased patients, changes in hospital policies, etc.) but also the interrelationships among
them.  The influence of some of the measured confounding variables was illustrated in the preceding sections. Section \ref{sec:GLM} shows the appropriate methodology for analysing data on counts of deaths when DP is on duty,  in the presence of possible confounding factors.

\section{GLM model}
\label{sec:GLM}
 Here we show how the results of a generalised linear model are in contrast to the findings in the report by the prosecution ``expert'' witnesses TM. 
TM's analysis is mainly focussed on the mortality rates of  nurses operating in the Lugo hospital. As already claimed within the introduction the two experts divide the hospital ward in two \textit{zones} made up of the two contiguous sectors, specifically sectors A + B and sectors C + D, calling the one \textit{same} zone -- if a nurse is on duty in either sector -- and    the other  \textit{opposite} zone. Then, for each shift, they compute the number of deaths recorded when a given nurse is on duty in a given zone and the number of deaths recorded in the opposite zone. As  DP has the highest patient mortality rate in the zone where she is on duty  compared to the mortality rate in the opposite zone  and thus the highest relative and absolute risk. In this way, TM aimed at inferring a causal effect between DP's presence  and the increase in deaths. For the sake of clarity we reproduce their computations in Table \ref{tab:micciolo} for  those nurses that work a similar number of hours as DP. First of all, we believe that introducing causation in this context is totally misleading. 

Causal effects cannot be inferred by just providing a descriptive analysis which can reveal, at most, an association among the presence of a given nurse and an increased mortality rate. Secondly, as already claimed in the previous sections, we believe that this association can be due to many confounding factors that if not accounted for appropriately can lead to a misleading conclusion.

 We now show that this association should simply be considered  spurious. Instead of focussing only on the deceased patients, we consider  all the patients hospitalized during the \textit{baseline} period. We create a new target variable,  a dummy variable called \textbf{\textit{Death}} which assumes value 1 if a given patient died during hospitalization and 0 otherwise. This allows us to model the probability of death in Lugo hospital as a function of many different covariates.  
 
 We build a logistic regression model where \textbf{\textit{Death}} is the response variable and the  covariates are:
 \begin{description}
 	\item[\textit{Sector}:] a categorical variable which represents the sector in which each patient was placed.
 	\item[\textit{Times}:]  a numerical variable that represents, for each patient, the number of shifts  when DP was on duty in the  same sector as the patient. This is obtained by considering the entire period of each patient's hospitalization and counting the number of shifts in which DP  was on duty in the same sector as the patient. For example, if a patient was hospitalized for 10 days, and during those 10 days,  DP worked  4 times in the same sector where the   patient was placed, then  \textit{Times} assumes value  4.
 	 	\item[\textit{Age}:] the patient's age. 
 	\item[\textit{Present}:] a dummy variable which assumes value 1 if a nurse works in the same zone  when a death is recorded and 0 otherwise.
 	(This  variable is used to reproduce TM's reasoning.)
 \end{description} 

The model was applied to two different data sets.  The first model (M1) uses   all the  admissions data. This model is estimated on all patients including those who were never under DP's care. Whereas, the second model (M2),  is estimated on a subset  of  the admissions data, \textit{i.e.} only those patients that were under DP's care at least once  are included in the analysis.

The results are given in Table \ref{tab:mod} which shows the coefficients, standard errors and significance of the variables included in models M1 and M2, as well as statistics for goodness-of-fit and model comparison. First, note that the two models are coherent both in terms of  coefficients' signs and statistical significance. Secondly,  observe that the variable associated with the sector where the patient is hospitalized is statistically significant for sectors B and  C. Patients hospitalized in these sectors have a lower  probability of death as their   coefficients are negative. This is in line with the findings  of the descriptive analysis  given in the previous sections. 
In fact, in Table \ref{tab:ricoveri}, we show that death rates in sectors B and C are  lower than those in A and D. Additionally, as expected,  \textit{Age}  is statistically significant and its  coefficient is positive, which even if trivial, confirms that older patients have higher probability of dying.
      
As far as DP is concerned the results from our analysis are totally in contrast to TM's conclusions. Note that the coefficient associated with covariate  \textit{Times} is negative. This means that, the more DP assists a patient, the lower the probability that the patient dies. This variable is statistically significant in both  models and, given the magnitude of its coefficient, it has a strong (negative) impact on the probability of death. Regarding  the presence of DP in the same zone where a death is recorded (variable \textit{Present}) we can definitely conclude that this has no relation at all with the probability that a patients dies. In fact, though  its estimated coefficient is positive, its standard error is extremely large: \textit{Present} is not statistically significant in any of the models and the $p$-value of its estimated coefficient is almost 1 (not shown).

In conclusion, our statistical models reveal that the increase in the number of deaths has no relation with DP's presence. The increase in number of deaths is a consequence both of random fluctuations and of the presence of those measured confounding factors which have, in part,  been analysed in these models.

\begin{table}[!hbtp] 
	\centering 
		\begin{tabular}{l|rr}
		 & Model M1 & Model M2 \\
		 \hline
		 Variable & Estimate & Estimate \\ 
		          &(std. err.) &(std. err.) \\
		          \hline
		Sector B  &  $-0.351^{**}$  &  $-0.870^{*}$ \\  
		&      (0.140) &  (0.484) \\  
		&            &        \\  
		Sector C  &  $-0.330^{**}$  &  $-1.058^{**}$ \\  
		&     (0.157) &  (0.435) \\  
		&            &        \\  
		Sector D  &     $-0.206$ &  $-0.345$ \\  
		&     (0.149) &  (0.281) \\  
		&            &        \\  
		Times  &  $-0.194^{***}$  &  $-0.136^{***}$ \\ 
		&     (0.030) &  (0.046) \\  
		&            &        \\  
		Age  &  $0.048^{***}$  &  $0.050^{***}$ \\  
		&     (0.006) &  (0.013) \\ 
		&            &        \\ 
		Present  &      20.36 &  21.49 \\  
		&       (313) &  (584) \\  
		&            &        \\  
		Intercept  &  $-5.550^{***}$  &  $-5.913^{***}$ \\  
		&     (0.488) &  (1.168) \\ 
			\hline 
		N. observations  &  \multicolumn{1}{c}{3,885}  &  \multicolumn{1}{c}{1,293} \\  
				Log Likelihood  &  \multicolumn{1}{c}{-1,272}  &  \multicolumn{1}{c}{-255} \\ 
		Pseudo-$R^2$  &  \multicolumn{1}{c}{0.30}  &  \multicolumn{1}{c}{0.62} \\  
		Akaike Inf. Crit.  &  \multicolumn{1}{c}{2,558}  &  \multicolumn{1}{c}{524} \\ 
		\hline  		        
		\textit{Note:}   &  \multicolumn{2}{r}{$^{*}p<0.1$; $^{**}p<0.05$; $^{***}p<0.01$}   \\ 
		\hline       
	\end{tabular}  
		\caption{Results of  logistic models M1 and M2.} 
	\label{tab:mod} 
\end{table}

\section{Prediction Intervals for the  ``Palacio-Tagliaro'' model}
\label{sec:pred}
	
Studies by \citet{dror2021cognitive} showed that forensic pathologists’  determinations of causes of death can be influenced by contextual information. A forensic pathologist might be more likely to determine that a patient’s death was  due to homicide if aware that the patient was under the care of a suspected nurse.   Even if he tries to ignore contextual information,  it may still bias the evaluation even without the pathologist even being aware of it. 

{The pathologist Tagliaro called as an expert witness for the prosecution in the first trial that lead to the life sentence conviction  for DP on 11/3/2016  based his prediction of the amount of  potassium ion $K^+$ in the  vitreous fluid of the eye of the deceased Rosa Calderoni on the simple linear regression of $K^+$ versus post-mortem interval (PMI) in Figure 2 of } 
\citet{bartolotti}.  

The sample used for estimating this regression line is based on  $n=67$ cases (40 males and 27 females) with a percentage of  40\% females and in only 3 cases do the females have an age over 55 years (i.e. 56, 58 and 63 years). The other females ranged in age from 8 to 53 years. Calderoni was 78 years old.  Thus, the sample is not representative of patients over 60. This factor is important as age has an influence on $K^+$ in vitreous humor \citep{zilg2015new}. The $K^+$ and PMI  in \cite{bartolotti} is not available so we were not able to reproduce a prediction interval on this data. Ironically, on  page 7 of \citet{pigaiani2020vitreous} (Tagliaro is  a co-author) they state:  \textit{For practical casework, literature therefore show that the
vitreous $K^+$ analysis cannot be used as a reliable test for the
diagnosis of fatal $K^+$ intoxication, given the rapid vitreous $K^+$
increase especially in the early PMI}.

A point prediction from a simple linear regression is useless. The prediction Tagliaro makes does not take into account any uncertainty or variability. Using a point prediction based on a linear regression that does not account for variability is extremely dangerous.
Furthermore, Tagliaro's work does not take into account the inter-individual variability due to several factors such as: the variability of the clinical condition, the conditions under which a death occurred and the preservation/temperature of the body, the variability due to the age of the deceased, the sex, etc. 

In a more recent paper in which Tagliaro is co-author  \cite{palacio2021simultaneous} 33 cases of violent or sudden death are examined. 
In their analysis they do not take into account inter-individual variability. In fact, it can be seen in their Table 1 that at similar PMIs, $K^+$ is highly variable. For example, for PMI values around 48 hours the range of $K^+$ varies between 13.6 and 20.6 mM (millimole). 
The authors themselves indicate the variability of the predictions, reporting that the errors of the estimate (difference between observed and estimated) are high. However in making the predictions presented in court, Tagliaro did not take this into account at all.

\subsection{Prediction Interval}
\label{sec:predint}

 Tagliaro measured the amount potassium ion $K^+$ in RC's vitreous humour 56 hours after her death.  Here we implement the regression model proposed in  \citet{palacio2021simultaneous} and calculate the prediction interval for  $K^+$ at PMI = 56 hours. The model is linear with a quadratic term, the dependent variable $Y$ is the concentration of $K^+$ measured in mM,  and the predictor $x$ is the PMI measured in hours.
Based on  the data in Table 1 page 4 of \citet{palacio2021simultaneous} the estimated model is 

\begin{equation*}
K^+=6.173+0.0127 PMI -0.0005 PMI^2
\end{equation*}

\begin{figure}[hbt!]
	\centering
	\includegraphics[scale=.75]{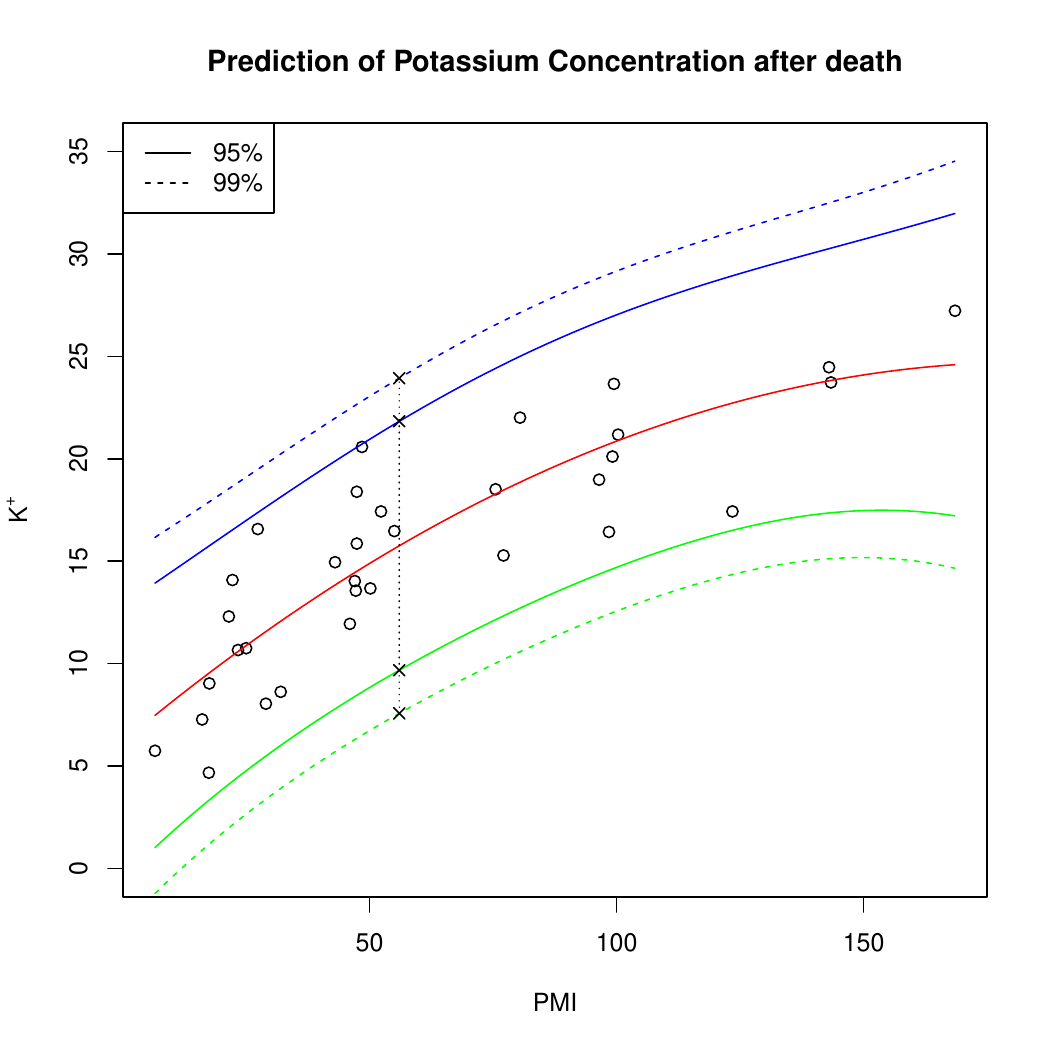}
	\caption{Regression model based on data in \citet{palacio2021simultaneous}  showing the 95\% and 99\% prediction curves and the corresponding interval at a PMI of 56 hours}.
	\label{fig:pred}
\end{figure}

\begin{table}[hbt!]
	\centering
	\begin{tabular}{rllll}
		\hline
		& Level & Point Estimate & Lower bound & Upper bound \\ 
		\hline
		& 95\% & 15.75 & 9.67 & 21.83 \\ 
		& 99\% & 15.75 & 7.57 & 23.94 \\ 
		\hline
	\end{tabular}
	\caption{Point estimate and prediction intervals for  $K^+$ at $PMI=56$}
	\label{tab:pred}
\end{table}

Figure \ref{fig:pred} shows the regression model based the  \citet{palacio2021simultaneous} data with the 95\% and 99\% prediction curves and highlights the corresponding interval at a PMI of 56 hours. Table \ref{tab:pred}  gives the values of the 95\% and 99\% prediction intervals when PMI=56 hours. The $K^+$ value of 19 mM that was taken on RC at PMI = 56 hours is well within the 95\% and 99\% prediction intervals [9.7,  21.8]  and [7.6, 23.9], respectively. This implies that the conclusions reached by Tagliaro are incorrect. The potassium value is in no way incriminating for DP.

\section{Conclusions}
\label{sec:conc}

The report of Tagliaro and Micciolo claims that deaths at Lugo hospital tended to occur more often
on days when DP was on duty than on days when she was not on duty. Their analysis based on ``days''  is statistically flawed for many reasons. A death registered to have occurred on a particular day is associated with a nurse even if she is on duty only for a fraction of that day (each shift lasts on average 8 hours). Also, nurses on night shifts were taken to be ``on duty'' on two consecutive days. Perhaps aware that a ``per day'' analysis could be very misleading, TM also looked at registered times of death. They claimed that deaths tended to occur more often on hours when DP was working than on hours when she was not working. This also leads to bias due to the mismatch between the actual time of death and time of registration of death. Deaths are recorded more often at 7 am (in the overlap between night and morning shift), and at midnight, and on the hour or on the half hour. 

The data shows that deaths occur more often in December and that the death to admissions rate fluctuates between different sectors over the years. There could be administrative decisions that concentrate more or less serious patients in the different sectors at different times. These could be some of the unmeasured confounders that might give the nurses working in these sectors a higher death rate than others.  
	
Measured and measurable confounders like the time of year, the time of day, the day of the week, the number of admissions, need to be controlled for, as they can have  a major effect on mortality rates, which may also have an association with the presence or absence of a given nurse. A difference in the mortality rates between different nurses could be due to those confounding variables. Even if all the measured confounders are taken into account, in an observational analysis like this no causal effect can be concluded, as there can be further unmeasured and unmeasurable confounding variables. For example, there may be changes in operating characteristics of the hospital not known to investigators or to outside experts or not thought to be important by hospital managers.
There could have been changes in hospital policy concerning policy on admissions, discharges and transfers to other wards, nursing homes, hospitals. There could have been changes in drug policy for the relief of pain and suffering in terminally ill patients, but that might increase the immediate risk of patients dying.
	
Each nurse has his or her own characteristics and different nurses can be assigned to different wards and different shifts partly through management decisions and partly through their own choices. Some nurses are thrilled to work extra hours, especially if there are difficult events taking place; others prefer to avoid such periods. Some prefer night shifts, some might prefer to avoid them. Different nurses can experience very different mortality rates for innocent reasons, some of which are not known or observable or measurable. Even if we could statistically demonstrate that the shift mortality rate of one particular nurse is, for example, 10\% higher than that of another (after taking into account measured confounders such as period of year, time of day, weekend, midweek day), it does not mean that there is a sinister reason for this. The reason could be entirely positive: for instance, that the most capable nurses work longer and harder shifts.

In conclusion we would like to refer  to a forthcoming  RSS report (under embargo) ``Healthcare Serial Killer or Coincidence?
Statistical Issues in Investigation of Suspected Medical Misconduct
RSS Statistics and the Law section'' to appear in September
2022. It is  specifically written  as a guideline for all people  (statisticians, judges, jury,  lawyers, pathologists  etc.)  and lay people 
working or reporting  on similar cases as this one.

\printbibliography  
\end{document}